\documentclass[preprint,aps]{revtex4-1}
\pdfoutput=1

\usepackage{amsmath,amssymb,amsfonts,dcolumn,color,graphicx,graphics,latexsym,placeins,epsfig}
\usepackage{epsfig}
\usepackage{bm}
\usepackage{slashed}
\usepackage{latexsym}
\usepackage{natbib}
\usepackage{url}
\usepackage{dcolumn}
\usepackage{color}
\usepackage{amsfonts,amssymb,amsmath}
\usepackage{graphicx,epsfig}
\usepackage{psfrag}
\usepackage{subfigure}
\usepackage{tabularx}
\usepackage{hyperref}
\hypersetup{colorlinks=true}

\newcommand{\be}{\begin{equation}}
\newcommand{\ee}{\end{equation}}
\newcommand{\ba}{\begin{eqnarray}}
\newcommand{\ea}{\end{eqnarray}}

\begin{document}

\title{Constraints on ultralight axions, vector gauge bosons, and unparticles from geodetic and frame-dragging measurements}
\author{Tanmay Kumar Poddar}
\email[Email Address: ]{tanmay@prl.res.in, tanmay.poddar@tifr.res.in}
\affiliation {Theoretical Physics Division, Physical Research Laboratory, Ahmedabad 380009, India}
\affiliation{Discipline of Physics, Indian Institute of Technology, Gandhinagar - 382355, India}
\affiliation{Department of Theoretical Physics, Tata Institute of Fundamental Research, Mumbai - 400005, India}


\begin{abstract}
The geodetic and frame-dragging effects are the direct consequences of the spacetime curvature near Earth which can be probed from the Gravity probe B (GP-B) satellite. The satellite result matches quite well with Einstein's general relativistic result. The gyroscope of the satellite which measures the spacetime curvature near Earth contains lots of electrons and nucleons. Ultralight axions, vector gauge bosons, and unparticles can interact with these electrons and nucleons through different spin-dependent and spin-independent operators and change the drift rate of the gyroscope. Some of these ultralight particles can either behave as a long range force between some dark sector or Earth and the gyroscope or they can behave as a background oscillating dark matter fields or both. These ultralight particles can contribute an additional precession of the gyroscopes, limited to be no larger than the uncertainty in the GP-B measurements. Compared with the experimental results, we obtain bounds on different operator couplings. 

\end{abstract}

\maketitle
\section{Introduction}
\label{sec1}
The gravity probe B (GP-B) is a satellite-based experiment that was launched in April 2004 to test the general relativity (GR) phenomena such as the geodetic effect or the de-Sitter effect \cite{deSitter:1916zz} and the Lense-Thirring precession \cite{Mashhoon:1984fj} or the gravitomagnetic frame-dragging effect predicted by Einstein's GR theory. This Earth satellite contained four gyroscopes and it was orbiting at $650~\rm{km}$ altitude. The spacetime near the Earth is changed due to the presence of the Earth and its rotation which modifies the stress-energy tensor near the Earth \cite{Schiff:1960gh}. When the gyroscope's axis of the satellite is parallel transported around the Earth then after one complete revolution, the tip does not end up pointing exactly in the same direction as before. The geodetic effect measures the drift rate of the gyroscope due to the presence of the Earth which wraps the spacetime near it and the frame-dragging effect measures the drift rate due to the rotation of the Earth which drags the spacetime around it. The GR predicts the geodetic drift rate as $-6606.1\hspace{0.1cm}~\rm{mas/yr}$ whereas the frame-dragging drift rate is $-39.2~\rm{mas/yr}$. In 2011, the GP-B experiment published the data for the geodetic drift rate as $-6601.8\pm 18.3~\rm{mas/yr}$ and the frame-dragging drift rate as $-37.2\pm 7.2~\rm{mas/yr}$, where $\rm{``mas"}$ stands for milliarcsecond \cite{Everitt:2011hp}. The GP-B result matches quite well with the GR predicted value. If there are interactions of electrons and nucleons in the gyroscope with the ultralight axion, gauge bosons, and unparticles through spin dependent and spin independent couplings then these interactions exert a new force on the gyroscope. This new force can contribute an additional precession of the gyroscopes. However, the contribution is within the measurements of experimental uncertainty. Comparing with the experimental results we obtain the constraints on axions, gauge bosons, and unparticles mediated forces. In the following, we have discussed what are axions and axion like particles, gauge bosons, and unparticles.

Axion is a pseudo-Nambu Goldstone Boson (pNGB) which was first postulated by Peccei and Quinn (1977) to solve the strong CP problem \cite{Peccei:1977ur,Peccei:1977hh,Weinberg:1977ma,Wilczek:1977pj}. The most stringent probe of the strong CP problem is the measurement of neutron electric dipole moment (nEDM). From chiral perturbation theory, one can obtain the nEDM as $d_n\simeq10^{-16}\bar{\theta}~\rm{e.cm}$, where $\bar{\theta}$ is related with the QCD $\theta$ angle by $\bar{\theta}=\theta+\rm{arg}\rm{det}(M)$ \cite{Adler:1969gk,Bell:1969ts}, where $\rm{M}$ is the general complex quark mass matrix. A natural choice of $\bar{\theta}\sim\mathcal{O}(1)$ violates the current experimental bound on nEDM which is $d^{expt}_n<10^{-26}~\rm{e.cm}$. Hence, $\bar{\theta}$ is as small as $\bar{\theta}<10^{-10}$ \cite{Baker:2006ts}. The smallness of $\bar{\theta}$ is called the strong CP problem. Quantum Chromodynamics (QCD) is the theory that can calculate the nEDM and its Lagrangian is 
\begin{equation}
\mathcal{L}=-\frac{1}{4}G^a_{\mu\nu}G^{a\mu\nu}+\sum^n_{j=1}[\bar{q}_ji\slashed{D}q_j-(m_j{q^\dagger_{Lj}} q_{Rj}+h.c.)]+\theta\frac{g^2_s}{32\pi^2}G^a_{\mu\nu}\tilde{G}^{a\mu\nu},
\label{i1}
\end{equation}
where $q$ denotes the quark field, $D$ denotes the covariant derivative, $\tilde{G}^{\mu\nu}=\frac{1}{2}\epsilon^{\mu\nu\alpha\beta}G_{\alpha\beta}$ is the dual gluon field tensor, $\theta$ is the QCD theta angle and h.c. denotes the Hermitian conjugate. The last term in Eq.\ref{i1} is a CP violating term. The last term comes from the symmetry of the Lagrangian and it must be present since all the quark masses are non-zero. However, QCD is good CP symmetric. To solve this strong CP problem, Peccei and Quinn came up with an idea that $\bar{\theta}$ is not just a parameter but a dynamical field that goes to zero by its classical potential. They postulated a global $U(1)_{PQ}$ symmetry which spontaneously breaks at a symmetry breaking scale called the axion decay constant $f_a$ and explicitly breaks due to non-perturbative QCD effects at a scale $\Lambda_{QCD}$, where the pseudo-Nambu Goldstone bosons called the QCD axions to get mass. The mass of the QCD axion is related to the axion decay constant as $m_a\simeq 5.7\times 10^{-12}\rm{eV}\Big(\frac{10^{18}\rm{GeV}}{f_a}\Big)$ \cite{GrillidiCortona:2015jxo}. Also, there exist other pseudoscalar particles which are not exactly the QCD axions but have similar kinds of interactions to the QCD axions. These are called Axion Like Particles or ALPs. These particles are motivated by string/M theory \cite{Witten:1984dg,Svrcek:2006yi,Conlon:2006tq,Arvanitaki:2009fg}. We assume the ALPs do not get any instanton induced mass and remain naturally light. For ALPs, $m_a$ and $f_a$ are independent of each other whereas, for QCD axions, $m_\pi f_\pi\sim m_af_a$, where $m_\pi$ is the pion mass and $f_\pi$ is the pion decay constant. The Lagrangian which describes the interaction between the ALPs and the standard model (SM) particles is \cite{Profumo:2019ujg}
\begin{equation}
\mathcal{L}\supset\frac{1}{2}\partial_\mu a\partial ^\mu a-\frac{\alpha_s}{8\pi}g_{ag}\frac{a}{f_a}G^{a}_{\mu\nu}\tilde{G}^{a\mu\nu}-\frac{\alpha}{8\pi}g_{a\gamma}\frac{a}{f_a}F_{\mu\nu}\tilde{F}^{\mu\nu}-\frac{1}{2f_a}g_{af}\partial_\mu a\bar{f}\gamma^\mu\gamma_5 f,
\label{i2}
\end{equation}
where $g$'s are the coupling constants. The first term denotes the kinetic term of ALP, the second term denotes the coupling of ALP with the gluon field $G_{\mu\nu}$, the third term denotes the coupling of ALP with the electromagnetic photon field $F_{\mu\nu}$ and the fourth term denotes the derivative coupling of ALP with the fermion field $f$. Note, $g_{ag}\ll 1$ to avoid any instanton induced mass for generic ALPs. The generic ALP is a pNGB, however, the scale at which the ALP gets mass need not be $\Lambda_{QCD}$. If the associated symmetry corresponding to a gauge group has mixed anomaly and has strong dynamics then the Goldstone boson acquires a mass $\sim \frac{\Lambda^2_a}{f_a}$, where $\Lambda_a$ is the scale where the new gauge group becomes strong. A small breaking of the global symmetry can also give the Goldstone boson a mass. This mass is proportional to the scale at which the global symmetry breaks. An example of breaking such global symmetry is quantum gravity. Here, the mass of the Goldstone boson is proportional to the breaking scale of the global symmetry. This leads to a tiny mass of ALPs. Thus, pNGB which does not get mass from the QCD instanton effects is generically called ALPs. The ALPs can also be generated from string theory. They arise as Kaluza-Klein zero modes of antisymmetric tensor fields in ten dimensions \cite{Witten:1984dg,Arvanitaki:2009fg,Acharya:2010zx}. The string axions inherit the shift symmetry from gauge invariance. They can also be generated from the clockwork mechanism \cite{Im:2019cnl}. Hence, for the QCD axion, the mass only depends on one free parameter $f_a$ whereas, for ALPs, the mass depends on additional parameters of the theory. The couplings of ALPs with SM particles are small since all the couplings are proportional to $\frac{1}{f_a}$ and $f_a$ generally takes a larger value. The axions or ALPs can also couple with the nucleons or quarks through the electric and magnetic dipole moment operators described by the terms $g_{EDM} a\bar{N}\sigma_{\mu\nu}\gamma_5 NF^{\mu\nu}$ and $g_{MDM} a\bar{N}\sigma_{\mu\nu}NF^{\mu\nu}$ respectively.

The axion is a promising non thermal dark matter candidate which can solve some of the small scale structure problems in the universe \cite{Nakagawa:2020eeg,Kitano:2021fdl,Arvanitaki:2019rax,Duffy:2009ig}. The axion field can oscillate with time as $a(t)\sim \frac{\sqrt{2\rho_{DM}}}{m_a}\sin(m_at)$, where $\rho_{DM}$ is the dark matter energy density. Axion can also form topological defects like cosmic strings and domain walls \cite{Kibble:1976sj,Kibble:1980mv,Takahashi:2020tqv}. They can also behave as dark radiation \cite{Cicoli:2012aq,Higaki:2012ar,Conlon:2013isa,Hebecker:2014gka,Dror:2021nyr,Jaeckel:2021gah}. If the mass of the axion is very small then it can also mediate long range forces and the corresponding potential is Yukawa type $\frac{1}{r}e^{-m_ar}$. The axions can also contribute to the monopole-monopole, monopole-dipole, and dipole-dipole interactions between the visible sector particles \cite{Moody:1984ba,Pospelov:1997uv,Adelberger:2009zz,Safronova:2017xyt}.

There is no direct evidence of axions so far. However, there are bounds on the axion mass and decay constant from the laboratory, astrophysical, and cosmological experiments. SN1987A puts the bound on axion decay constant as $f_a\gtrsim 10^9~\rm{GeV}$. The CAST experiment puts bounds on the solar axion \cite{CAST:2017uph,Inoue:2008zp,Arik:2008mq}. Axions can be a component of hot dark matter if $f_a\lesssim 10^8~\rm{GeV}$ \cite{Hannestad:2005df,Melchiorri:2007cd,Hannestad:2008js}. Cold axions can be produced from vacuum realignment mechanism which can give cold dark matter relic density \cite{Hertzberg:2008wr,Visinelli:2009zm}. Some laboratory and cosmological bounds on axions are discussed in \cite{Semertzidis:1990qc,Cameron:1993mr,Robilliard:2007bq,Chou:2007zzc,Sikivie:2007qm,Kim:1986ax,Cheng:1987gp,Rosenberg:2000wb}. Axions can also be probed from the superradiance phenomena for a black hole \cite{Plascencia:2017kca,Chen:2019fsq,Arvanitaki:2010sy,Arvanitaki:2014wva,Brito:2015oca,Day:2019bbh}. Ultralight axion like particles can be a candidate of fuzzy dark matter (FDM) if $m_a\sim 10^{-22}~\rm{eV}$ and $f_a\sim 10^{17}~\rm{GeV}$ \cite{Hu:2000ke,Hui:2016ltb}. The de Broglie wavelength of the FDM particles is of the order of the size of a dwarf galaxy $(1-2)~\rm{kpc}$. However, Lyman-$\alpha$ bounds disfavor ALPs as FDM \cite{Kobayashi:2017jcf,DES:2020fxi}. The Shapiro time delay also puts constraints on the ALPs as $m_a\lesssim 10^{-18}~\rm{eV}$ and $f_a\lesssim 10^7~\rm{GeV}$ \cite{Poddar:2021sbc}. The neutron star-white dwarf binary systems put bounds on ALPs as $m_a\lesssim 10^{-19}~\rm{eV}$ and $f_a\lesssim 10^{11}~\rm{GeV}$ \cite{Poddar:2019zoe}. The axions can also be constrained from the birefringence phenomena due to the interactions of axions and electromagnetic photons \cite{Sigl:2018fba,DeRocco:2018jwe,Fedderke:2019ajk,Agrawal:2019lkr,BICEPKeck:2020hhe,Poddar:2020qft,BICEPKeck:2021sbt}. There are few experiments like ABRACADABRA \cite{Kahn:2016aff,Ouellet:2018beu}, CASPEr \cite{Graham:2013gfa,Budker:2013hfa,JacksonKimball:2017elr,Garcon:2017ixh}, GNOME \cite{Pustelny:2013rza,Afach:2018eze,Afach:2021pfd} which are looking for axions using magnetometers \cite{Kim:2021eye}. Storage rings can also be used for the detection of axions \cite{Graham:2020kai}. Constraints on axion mediated force from torsion pendulum experiment are discussed in \cite{Hoedl:2011zz}.

The light gauge bosons can also mediate long range forces or can also serve as a background oscillating dark matter fields. The SM of particle physics is a $SU(3)_c\times SU(2)_L\times U(1)_Y$ gauge theory. However, in the leptonic sector one can construct three symmetries, $L_e-L_\mu$, $L_e-L_\tau$, and $L_\mu-L_\tau$ in an anomaly free way and they can be gauged. The $L_e-L_{\mu,\tau}$ type of gauge force can be constrained from neutrino oscillation experiments \cite{Joshipura:2003jh}, and perihelion precession of planets \cite{Poddar:2020exe}. The $L_\mu-L_\tau$ type of gauge force can be constrained from the orbital period loss of the binary systems \cite{Poddar:2019wvu}. The other bounds on $L_i-L_j$ force are discussed in \cite{Dror:2020fbh,Coloma:2020gfv}. $B-L$ symmetry can also be gauged in an anomaly free way and it can mediate long range force. The bounds on ultralight $B-L$ gauge bosons are discussed in \cite{LopezNacir:2018epg}. Constraints on long and short range forces mediated by scalars and vectors are discussed in \cite{Vasilakis:2008yn,Yan:2014yva}. The dark photon which is another standard dark matter candidate can have a kinetic mixing with the visible photons or couple with the SM particles if they are charged under the new $U(1)$ gauge group. Its mass can be light and mediate long range force as usual \cite{Banerjee:2019pds,Gaidau:2021vyr,Lu:2021uec,Tsai:2021irw}. 

Unparticles are also good candidates that can mediate long range forces. In 2007, Georgi proposed a new type of massless particles that can arise in effective low energy theory due to its non canonical dimensions \cite{Georgi:2007ek,Georgi:2007si}. Due to its non canonical scaling dimensions, the force mediates by unparticle can deviate from inverse square law and can give rise to long range forces. Suppose, an ultraviolet (UV) theory has an infrared (IR) fixed point at some energy scale $\Lambda_\mathfrak{u}$. The fields become conformally invariant at this scale. If the operator of the UV theory is denoted by $\mathcal{O}_{UV}$ with a canonical dimension $d_{UV}$ and if any SM operator is denoted by $\mathcal{O}_{SM}$ with dimension $d_{SM}$, then the effective coupling is suppressed by some heavy mass scale $M_\mathfrak{u}$ and is denoted by $\frac{1}{M^{d_{UV}+d_{SM}-4}_\mathfrak{u}}$. The fields of the UV theory become scale invariant below a scale $\Lambda_\mathfrak{u}$ (typically $\Lambda_\mathfrak{u}\sim 1 \hspace{0.1cm}\textrm{TeV}$) and it acquires a dimension $d_\mathfrak{u}$ different from the canonical one by the dimensional transmutation. Thus, the unparticle operator $O_\mathfrak{u}$ coupled with the SM operator as $\Big(\frac{\Lambda_\mathfrak{u}}{M_\mathfrak{u}}\Big)^{d_{UV}+d_{SM}-4}\frac{1}{\Lambda^{d_\mathfrak{u}+d_{SM}-4}_\mathfrak{u}}\mathcal{O}_\mathfrak{u}\mathcal{O}_{SM}$. 

Exchange of scalar, pseudoscalar, vector, pseudovector unparticles can give rise to long range forces which are discussed in \cite{Liao:2007ic,Deshpande:2007mf,Adelberger:2006dh,Hunter:2013hza}. Unparticles can couple to energy stress tensor and mediate ungravity \cite{Goldberg:2007tt}. The unparticle coupling with Higgs, gauge bosons, and other SM particles are discussed in \cite{Kikuchi:2007qd,Chen:2007vv,Lu:2007mx,Greiner:2007hr,Davoudiasl:2007jr,Luo:2007bq,Neubert:2007kh,Bhattacharyya:2007pi,Delgado:2008gj}. Unparticle mediated long range force can also be tested from the perihelion precession of Mercury \cite{Das:2007cc}. Also, dark matter and dark energy can interact with unparticles which are discussed in \cite{Kikuchi:2007az,Chen:2009ui,Dai:2009mq,Jamil:2011iu,Abchouyeh:2021wey}.

Axions, vector gauge bosons, and unparticles can mediate long range forces and contribute to the precessional velocity of the gyroscope of GP-B satellite within the experimental uncertainty limit. The axions, dark photons, vector gauge bosons, and unparticles can interact with the gyroscope and change the precession of the gyroscope. Using the GP-B results, we obtain bounds on the coupling and mass of these particles.

The paper is organized as follows. In Sec.\ref{new} we calculate the fraction of polarized spins in the GP-B gyroscope to constrain spin dependent coupling. In Sec.\ref{sec2}, we consider axion mediated long range Yukawa type of potential between a visible sector and a dark sector. In Sec.\ref{sec3}, we discuss the interaction of the GP-B gyroscope with the background oscillating axionic dark matter field. In Sec.\ref{sec4}, we consider that the time dependent oscillating axionic field can interact with the nucleons of the gyroscope through the electric dipole moment operator and change its drift rate. The dark photon can also behave as a background oscillating dark matter field and can interact with the gyroscope's nucleons through electric and magnetic dipole moment operators which are discussed in Sec.\ref{sec6}. In Sec.\ref{sec5}, we discuss the mediation of $L_e-L_{\mu,\tau}$ type of gauge bosons which gives rise to long range force and changes the precession rate of the gyroscope. In Sec \ref{sec7}, we discuss the unparticle mediated long range force which can be constrained from the GP-B result. Finally, in Sec.\ref{sec8}, we discuss our results.

We have used the natural system of units throughout the paper.
\section{Estimation of the fraction of polarized spins in GP-B gyroscope}\label{new}
Unless any spin alignment mechanism, macroscopic objects with randomly oriented spins would not induce any spin dependent force. In \cite{Hunter:2013hza}, Earth has been treated as a polarized electron source to search for long range spin-spin interaction. A fraction of the total number of electrons in Earth can be polarized due to the presence of Earth's dipolar magnetic field. However, for the measurements of frame-dragging and geodetic effects, GP-B satellite explicitly shielded external magnetic fields (e.g., the Earth field) to $<0.1~\rm{nT}$. If there is no shielding in GP-B, the spin alignment would presumably have to adiabatically track the Earth's field direction as the satellite orbited in its polar orbit through the Earth's mostly dipolar (with higher moments) field. This alone would place a torque on the gyroscopes that, if a large number of polarized spins were achieved, may have been problematic from the perspective of the GR mission. The precession of the gyroscope would then take contributions both from spacetime curvature as well as background magnetic field. However, the quartz gyroscopes are coated with superconducting niobium. When the niobium coated gyroscope rotates, it can produce a small magnetic field. The London moment induced field for the GP-B gyros is likely on the order of $B_{\rm{London}}\sim \Big(\frac{2M}{Q}\Big)\omega=1.08\times 10^{-8}~\rm{T}$, assuming $M\sim 2m_e$, $Q\sim 2e$, and with $\omega$ being the angular frequency corresponding to the $\sim 9000~\rm{rpm}$ gyro spin rate achieved in GP-B. Hence, the London-moment field is the larger possible field and it is aligned with the gyro rotational axis. The quartz spheres that form the GP-B gyros are silica $(\rm{SiO_4})$ which consists of $^{14}\rm{Si}$ and $^{17}\rm{O}$ isotopes. The nuclear moments are respectively, $\mu_{^{14}\rm{Si}}\sim 0.6\rm{\mu_N}$ and $\mu_{^{17}\rm{O}}\sim 1.9\rm{\mu_N}$, where $\mu_N=(e/2m_p)$ is the nuclear magneton ($m_p$ is the mass of proton). Here, we consider the oxygen spins to be the relevant ones, both because there is more oxygen than silicon in silica and because it has a larger moment. For the $T_B=2.5~\rm{K}\sim 2.15\times 10^{-4}~\rm{eV}$ temperature environment in GP-B, the energy splitting for the maximally spin-aligned and maximally anti-aligned states for the $^{17}\rm{O}$ nucleus is $\Delta E\sim 2 \times \mu_{^{17}O}B\ll kT_B$, so the estimate for spin polarization for the oxygen nuclei is of the order of $\alpha\sim 1-\exp[-\Delta E/(kT_B)]\sim \Delta E/(kT_B) \sim 2 \times \mu_{^{17}O}B/ kT_B\sim 5.6\times 10^{-12}$. There is some small correction $(\mathcal{O}(1)$ factors) for the presence of the $\rm{Si}$, but this is a rough order of magnitude estimate of spin polarization fraction due to London-moment induced magnetic field. The value of $\alpha$ is tiny, although larger than the accidental polarization estimate of $\frac{1}{\sqrt{N_{\rm{nucleons\hspace{0.1cm}in\hspace{0.1cm}gyro}}}}\sim 1.7\times 10^{-13}$. Note, the accidental net spin in absence of an alternative alignment mechanism would quite likely vary stochastically in both magnitude and direction over time on something like the spin coherence time, governed by how the randomly aligned spins in the quartz gyro interact. This would lead to an additional suppression of any new physics precession on the gyro, as the effect would execute a random walk instead of a coherent linear growth. This will cause an additional suppression by $\sim \sqrt{T_{\rm{spin}}/T}\ll 1$ (with $T_{\rm{spin}}$ the spin coherence time and $T$ the mission duration) and $\alpha_{\rm{accidental}}\ll \alpha$. In the following, we will constrain the spin dependent coupling due to the presence of London-moment induced magnetic field which is the larger possible field for which the fraction of polarized spins is $\alpha\sim 5.6\times 10^{-12}$. 
\section{Constraints on axion mediated long range Yukawa type of potential}
\label{sec2}
The generic ultralight axion like particles (ALPs) which arise in string/M theory has a shift symmetry $a\rightarrow a+\theta$ ($\theta$ is a real number) and transforms as $a\rightarrow -a$ under CP symmetry. Since CP is violated in nature, we consider CP violating coupling of axions with the dark sector. However, in the QCD sector, CP is conserved and we consider CP conserving coupling of axions with nucleons. Due to very light mass, the ultralight ALPs can mediate a long range Yukawa type of potential $(\sim \frac{1}{r}e^{-m_a r}$, $m_a$ is the mass of the ALP) between the visible sector which consists of standard model particles and the dark sector which consists of dark matter particles. The dark sector can also consist of some compact objects formed by axion dark matter, topological defects such as domain walls, cosmic strings or even it can be primordial black holes (PBH) \cite{Carr:1975qj,Hawking:1975vcx,Carr:2016drx}. The mass of the ALP is typically constrained by the distance between the dark sector and the visible sector. The range of the ALP mediated force is $\lambda\gtrsim\frac{1}{m_a}$ which makes the ALPs ultralight. The equation of motion of the ALP field $(a)$ from the CP violating coupling with the dark sector is \cite{Kim:2021eye}
\begin{equation}
\nabla_\mu\nabla^\mu a(t,\vec{x})=-m^2_a a(t,\vec{x})-J(t,\vec{x}),
\label{eq:1}
\end{equation}    
where the source term $J(t,\vec{x})$ denotes the current density in the CP violating dark sector. Such CP violating coupling of axions with the dark sector arises in a dark QCD model where the dark nucleon couples with the axion by a monopole force $(\mathcal{L}_{\rm{eff}}\supset g_{a N^\prime N^\prime}a \bar{N}^\prime N^\prime$, where $N^\prime$ is the dark nucleon) if the strong CP phase is not tuned \cite{Moody:1984ba,Pospelov:1997uv,Kim:2021eye}. The axion DM can also emit monopole force in the axiverse scenario with several axions and the vacuum of the axion potential is CP violating \cite{Acharya:2010zx,Marsh:2019bjr}. The axionic topological defects can also emit monopole force \cite{Kibble:1976sj,Kibble:1980mv,Takahashi:2020tqv}. Since, $J(t,\vec{x})$ is the current density, it should be proportional to the energy density of the CP violating dark sector, $\rho_D$. From the dimensional analysis, it is also inversely proportional to the energy scale in the theory, here it is $f_a$, the axion decay constant. Hence, one can parametrize the current density as $J(t,\vec{x})=c_D\kappa(t)\frac{\rho_D(\vec{x})}{f_a}$, where $c_D$ denotes the strength of the CP violation in the dark sector, and $\kappa(t)$ includes other model dependent parameters. In a Schwarzschild spacetime background, the Klein-Gordon equation (Eq.\ref{eq:1}) for the axion field cannot be exactly solvable for a point source with $J_0(\vec{x})=\delta^3(\vec{x})$. Hence, to solve Eq.\ref{eq:1}, we expand the axion field in a perturbative way with the perturbation parameter $\frac{GM_D}{R_D}$ (where $G$ denotes Newton's gravitation constant, $M_D$ denotes the mass of the dark object and $R_D$ denotes its radius) and keeping the leading order term with its Yukawa behaviour. Hence, we can write the homogeneous solution for the axion field as 
\begin{equation}
a_0(r)=a_1(r)+\frac{GM_D}{R_D}a_2(r)+\mathcal{O}\Big(\frac{GM_{D}}{R_D}\Big)^2,
\label{eq:2}
\end{equation}
where the leading order term $a_1(r)=\frac{1}{r}e^{-m_ar}$ has an Yukawa behaviour. Putting Eq.\ref{eq:2} in Eq.\ref{eq:1} in the Schwarzschild spacetime background, we obtain the resultant homogeneous axion field solution as 
\begin{equation}
a(r)=\frac{e^{-m_ar}}{r}\Big[1+\frac{GM_D}{r}\{1-m_ar\ln(m_ar)+m_are^{2m_ar}E_i(-2m_ar)\}\Big]+\mathcal{O}\Big(\Big(\frac{GM_D}{R_D}\Big)^2\Big),
\label{eq:3}
\end{equation}
where $E_i(x)=-\int^\infty_{-x}\frac{e^{-t}}{t}dt$ is called the exponential integral function. Since we are not considering any particular dark sector object therefore, we do not have any information about $M_D$ and $R_D$. Hence, we keep ourselves in the regime $\frac{GM_D}{R_D}\ll1$ and the axion field solution (basically, the Green's function) becomes $a_0(r)\approx\frac{1}{r}e^{-m_ar}$. Hence, the axion field for the stationary non relativistic dark source is
\begin{equation}
a(r)=\int d^3\vec{x}^\prime \frac{c_D\rho_D(\vec{x}^\prime)}{f_a}a_0(|\vec{x}-\vec{x}^\prime|),
\label{eq:4}
\end{equation}
where we have assumed $\kappa(t)=1$.

The derivative coupling of ALP with the nucleons in the CP conserving SM sector is governed by the Lagrangian
\begin{equation}
\mathcal{L}=\frac{c_{f}\partial_\mu a}{f_a}\bar{f}\gamma^\mu\gamma_5f,
\label{eq:5}
\end{equation}
where $c_{f}$ denotes the dimensionless constant for the SM fermions $(f=p,n)$. In the non-relativistic limit, the Dirac bilinears take the following form 
\begin{equation}
<\bar{f}\gamma^0f>\sim N_f, \hspace{0.2cm} <\bar{f}\vec{\gamma}f>\sim<v_f>, \hspace{0.2cm} <\bar{f}\gamma_0\gamma_5f>\sim <\frac{\sigma_fp_f}{E_f}>, \hspace{0.2cm} <\bar{f}\vec{\gamma}\gamma_5f>\sim<\sigma_f>.
\label{eq:6}
\end{equation}
Hence, from Eq.\ref{eq:5}, we can write the interacting Hamiltonian in the non relativistic limit of the fermions as
\begin{equation}
\mathcal{H}\approx -\frac{c_{f}}{f_a}\vec{\nabla} a.\vec{\sigma}_f.
\label{eq:7}
\end{equation}
Also, the Hamiltonian for a particle having spin in a magnetic field $(\vec{B})$ has a magnetic moment $(\mu_f\sigma_f)$ interaction as $\mathcal{H}^\prime=-\mu_f\vec{\sigma}_f.\vec{B}$. Comparing $\mathcal{H}^\prime$ with $\mathcal{H}$ in Eq.\ref{eq:7}, we obtain the induced magnetic field due to the mediation of long range axionic Yukawa field as 
\begin{equation}
\vec{B_a}=\frac{c_f\vec{\nabla}a}{\mu_ff_a},
\label{eq:8}
\end{equation}
where $\mu_n\approx -1.9\mu_0$, $\mu_p\approx 2.8\mu_0$ and $\mu_0=\frac{e}{2m_N}=1.5\times 10^{-10}~\rm{eV^{-1}}$ in natural units. Due to the primordial density fluctuations and the structure formations, the dark matter of the CP violating dark source is not spatially homogeneous and we have a non zero value of $B_a$. But if the dark matter distribution is spatially homogeneous then $B_a$ is zero even if $c_f\neq0$. The energy density of the dark matter $\rho_D$ in Eq.\ref{eq:4} can be decomposed into the galactic $(\rho_{gal})$ and the extragalactic parts $(\rho_{egal})$. We take the galactic contribution of dark matter $\rho_{gal}$ as simply the Navarro-Frenk-White (NFW) profile defined as \cite{Navarro:1995iw,Cirelli:2010xx}
\begin{equation}
\rho_{gal}(r)=\rho_{NFW}(r)=\frac{\rho_s r_s}{r}\Big(1+\frac{r}{r_s}\Big)^{-2},
\label{eq:9}
\end{equation} 
where $\rho_s=0.184~\rm{GeV/cm^3}=1.428\times 10^{-6}~\rm{eV^4}$, $r_\odot\approx8.33~\rm{kpc}=1.297\times 10^{27}~\rm{eV^{-1}}$, and $r_s=24.43~\rm{kpc}=3.807\times 10^{27}~\rm{eV^{-1}}$. Hence, around the galactic centre, the contribution from $\rho_{gal}$ is important and the axion induced magnetic field is approximately
\begin{equation}
\begin{split}
B_a^{gal}(r)\approx-\frac{c_fc_D\rho_s r_s}{\mu_f f^2_a}\Big[2\Big(E_i(-m_a r)-e^{r_sm_a}E_i(-m_a(r_s+r))\Big)+\\
\frac{r_se^{-m_a r}\Big(r_s+2r+2m_a e^{m_a(r_s+r)}(r_s+r)^2E_i(-m_a(r_s+r))\Big)}{(r_s+r)^2}\Big].
\end{split}
\label{eq:10}
\end{equation}
The extragalactic contribution becomes dominant for the region where the Compton wavelength of the ALP $(\frac{1}{m_a})$ is much larger than the size of the galaxy. The dark matter energy density from the extragalactic part within the horizon is $\rho_c\times \Omega_{DM}$, where $\rho_c=1.1\times 10^{-5}h^2~\rm{GeV}/\rm{cm^3}=3.83\times 10^{-11}~\rm{eV}^4$, is the critical density of the universe and $\Omega_{DM}h^2\simeq 0.12$ is the relic density of the dark matter. $h=0.67$ denotes the reduced Hubble parameter. Due to the primordial density fluctuations $(\mathcal{O}(10^{-5}))$ from inflation, an inhomogeneity exists in $\rho_{egal}$ which is parametrized as
\begin{equation}
\vec{\eta}=\Big(\frac{\rho_c\Omega_{DM}}{3m_a}\Big)^{-1}\int d^3\vec{x}^\prime \rho_{egal}(\vec{x}^\prime)\frac{\vec{x}-\vec{x}^\prime}{|\vec{x}-\vec{x}^\prime|^3}e^{-m_a|\vec{x}-\vec{x}^\prime|}.
\label{eq:11}
\end{equation}
Hence, the induced magnetic field mediated by long range axion force due to the extragalactic contribution of dark matter is 
\begin{equation}
B^{egal}_a(r)\approx\frac{c_Dc_f\eta\rho_{crit}\Omega_{DM}}{\mu_f f^2_am_a}e^{-m_ar}(2+m_a r).
\label{eq:12}
\end{equation}
The axion mediated force from the extragalactic contribution increases with decreasing the axion mass while keeping the decay constant $f_a$ fixed. When $m_a^{-1}\gg r_\odot$, the force dominates around the galactic centre. However, if $m_a^{-1}$ is much larger than the size of the galaxy then the force due to the extragalactic term becomes important. Though from numerical estimation, the density perturbation term is subdominant in the mass range that we are considering. If the mass of the axion is less than $10^{-33}~\rm{eV}$ which is constrained by the horizon size, then the axion mediated force could not reach us. The total induced magnetic field due to the long range axion mediated force is $B_a(r)=B_a^{gal}(r)+B_a^{egal}(r)$. The direction of the effective magnetic field towards or opposes the galactic centre. The quantity $|\mu_fB_a|$ governs an additional precession of GP-B gyroscope and the effect of axion mediated long range force is within the experimental uncertainty in the measurements of geodetic and frame-dragging effects. 
\begin{figure}
\centering
\includegraphics[width=4.0in,angle=360]{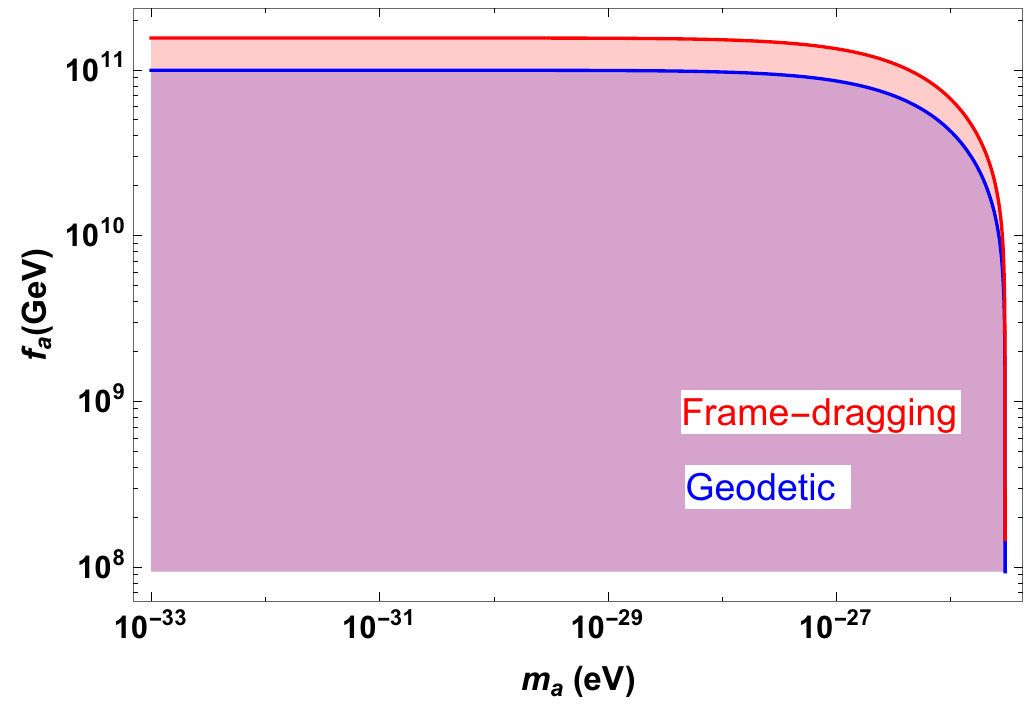}
\caption{Variation of $f_a$ with $m_a$ from geodetic (blue) and frame-dragging (red) measurements using GP-B result for a long range axion mediated force between a visible sector and a dark sector with fixed values of the parameters $\eta=10^{-5}, c_f=1,c_D=1$. The shaded regions below these lines are excluded.}
\label{fig:plot1}
\end{figure}
In Fig.\ref{fig:plot1} we have shown the variation of axion decay constant $(f_a)$ with the axion mass $(m_a)$ for the axion mediated force between the CP violating dark sector and the CP conserving SM sector from frame-dragging and geodetic measurements using the GP-B results. We obtain the bounds on the ALP decay constant by calculating $B_a$ numerically as $f_a\gtrsim 1.6\times 10^{11}~\rm{GeV}$ from frame-dragging effect (red line) and $f_a\gtrsim 1.0\times 10^{11}~\rm{GeV}$ from geodetic effect (blue line) for the axion mass $m_a\lesssim 10^{-26}~\rm{eV}$. The shaded regions below these lines are excluded. The frame-dragging effect puts a stronger bound on $f_a$. We have fixed $\eta=10^{-5}, c_f=1,c_D=1$ and the fraction of polarized spins $\alpha\sim 5.6\times 10^{-12}$ due to London-moment induced magnetic field to obtain these bounds. We have checked that the dominant contribution in the drift rate due to long range axion force comes from the galactic contribution. The axion induced effective magnetic field cannot be shielded like the ordinary magnetic field and it remains intact. The magnitude of this magnetic field remains constant within the experimental timescale and its direction towards or opposes the galactic centre as discussed above. Such pseudo magnetic field can be measured by SQUID (Superconducting QUantum Interference Device) within the quartz housing of the telescope \cite{2003AdSpR..32.1397M,2008PhDT........64S,Everitt:2015qri}.
\section{Constraints on oscillating axionic field from GP-B result}
\label{sec3}
The axionic field can also oscillate with time and behave as a background. The oscillating field can interact with the spin of the gyroscope of GP-B satellite and affects the precession rate. The oscillating axionic field can be defined as 
\begin{equation} 
a(t)=\frac{\sqrt{2\rho_{DM}}}{m_a}\sin(m_at),
\label{s1}
\end{equation}
where $\rho_{DM}=0.3~\rm{GeV/cm^3}=2.33\times 10^{-6}~\rm{eV^4}$ is the local dark matter density. The axions can have a derivative coupling with the nucleons as described by the Lagrangian Eq.\ref{eq:5}. In the rest frame of the gyroscope, the corresponding Hamiltonian is 
\begin{equation}
\mathcal{H}=-g_{aNN}\vec{\nabla} a.\vec{\sigma},
\label{s2}
\end{equation}
where $\sigma$ is the spin of the nucleon that precesses in presence of a magnetic field characterised by $\vec{\nabla} a$ and $g_{aNN}\propto \frac{1}{f_a}$ is the axion nucleon coupling in the unit of $\rm{energy}^{-1}$. Hence, using Eq.\ref{s1}, the time dependent effective magnetic field becomes 
\begin{equation}
B_a(t)=\frac{g_{aNN}}{\mu_N}\sqrt{2\rho_{DM}}v\cos(m_at),
\label{s3}
\end{equation}
where $\mu_N=-1.9\mu_0(2.8\mu_0)$ for neutron (proton), $\mu_0=\frac{e}{2m_N}\approx0.1~\rm{e.fm}$ is the nuclear magneton, and $v$ is the relative velocity of the gyroscope. This time dependent induced magnetic field can exert a force on the gyroscope as a whole and the precessional velocity of the gyroscope due to the oscillating axionic field becomes
\begin{equation}
\mu_NB_a(t)=g_{aNN}\sqrt{2\rho_{DM}}v\cos(m_at).
\label{s4}
\end{equation}
Eq.\ref{s4} can contribute to the precessional velocity of the gyroscope and its contribution is within the experimental uncertainties in the measurements of frame-dragging and geodetic effects.
\begin{figure}
\centering
\includegraphics[width=4.0in,angle=360]{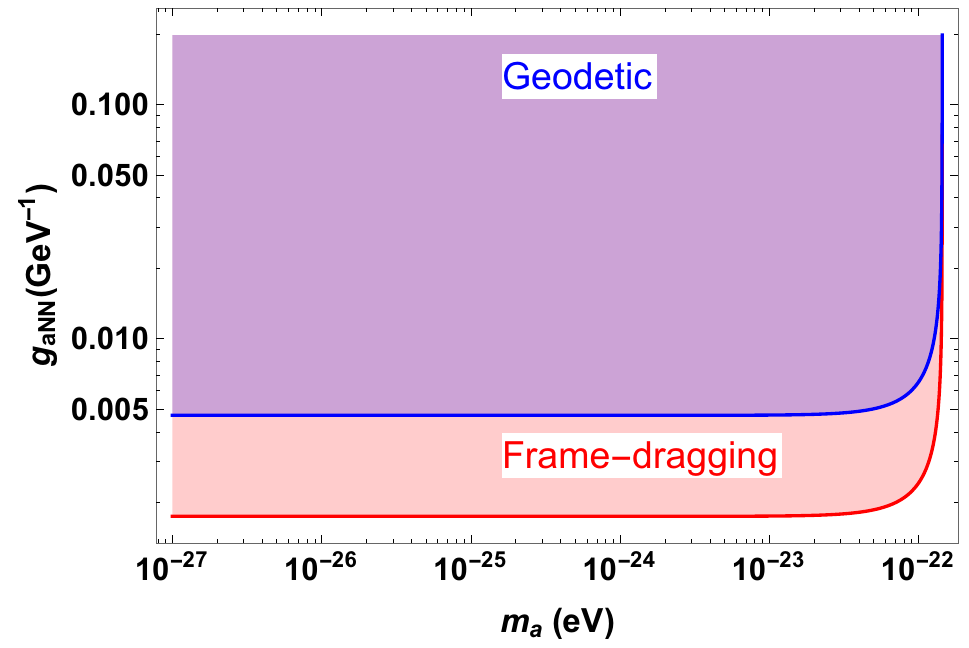}
\caption{Variation of $g_{aNN}$ with $m_a$ for the geodetic (blue) and the frame-dragging (red) measurements using GP-B result for time oscillating background axionic field. The shaded regions above these lines are excluded.}
\label{fig:plot3}
\end{figure}
In Fig.\ref{fig:plot3} we have shown the variation of $g_{aNN}$ with $m_a$ for the geodetic (blue) and the frame-dragging (red) measurements using GP-B result for time oscillating axionic field. The regions above these lines the excluded. We obtain the bounds on the axion-nucleon coupling as $ g_{aNN}\lesssim 4.8\times 10^{-3}~\rm{GeV^{-1}}$ from the geodetic measurement and $ g_{aNN}\lesssim 1.7\times 10^{-3}~\rm{GeV^{-1}}$ from the frame dragging measurement due to London moment induced magnetic field. The frame dragging gives the stronger bound on $g_{aNN}$. Here, $t=T=11~\rm{months}=4.33\times 10^{22}~\rm{eV^{-1}}$ denotes the GP-B data taking duration. If $m_aT\ll 2\pi$ then the torque on the gyroscope from the axion field is approximately a constant and the precession angle increases linearly with the mission duration. However, if $m_aT\gg 2\pi$, then the torque on the nucleon spin flips back and forth during the mission and one can only get a net precession effect from the last uncancelled period of the oscillation which means that the net precession is suppressed by $\sim \frac{2\pi}{m_a T}$ (or faster) as compared to the case of linear growth of the precession angle. The effect adds coherently for $m_a\lesssim \frac{2\pi}{T}$  whereas for $m_a\gtrsim \frac{2\pi}{T}$ the precessional velocity oscillates with time. Hence, the bounds on the coupling are only applicable for the axions of mass $m_a\lesssim \frac{2\pi}{T}\sim 1.45\times 10^{-22}~\rm{eV}$. 
\section{Constraints on axion EDM coupling from the GP-B result}
\label{sec4} 
The axions can couple with the nucleons through the electric dipole moment operator described by the Lagrangian 
\begin{equation}
\mathcal{L}\supset g_d a\bar{N}\sigma_{\mu\nu}\gamma_5 NF^{\mu\nu},
\label{l1}
\end{equation}
where $g_d$ is the coupling constant. Hence, in the non relativistic limit, the precessional velocity of the gyroscope due to the electric dipole moment operator becomes
\begin{equation}
v^d_{prec}(t)\sim g_da(t)E\sim g_d\frac{\sqrt{2\rho_{DM}}}{m_a}\sin(m_at)E,
\label{l2}
\end{equation}
where $a$ is the time oscillating axionic field, $E$ is the electric field induced by the Earth's magnetic field in the rest frame of the satellite, and it is $\vec{E}\sim \vec{v}\times \vec{B}$. We take the Earth's magnetic field as $B\sim 0.1~\rm{Gauss}$ and we assume that it is attenuated by a factor of $\delta\sim 10^{-6}$. Hence, the value of the electric field becomes $E=2.51\times 10^{-12}~\rm{Gauss}=4.89\times 10^{-14}~\rm{eV^2} (1~\rm{Gauss}=1.95\times 10^{-20}~\rm{GeV^2}$). Considering the precessional velocity due to the effect of axion EDM coupling is within the experimental uncertainties of the GP-B measurements, we can write the oscillation amplitude from Eq.\ref{l2} as 
\begin{equation}
g_d\lesssim 5.02\times 10^{3}\rm{GeV^{-2}}\Big(\frac{m_a}{eV}\Big),
\label{l3}
\end{equation}
for frame dragging effect and 
\begin{equation}
g_d\lesssim 1.35\times 10^{4}\rm{GeV^{-2}}\Big(\frac{m_a}{eV}\Big)
\label{l4}
\end{equation}
for the geodetic effect. The bounds on the EDM coupling are only applicable for the axions of mass $m_a\lesssim \frac{2\pi}{T}\sim 1.45\times 10^{-22}~\rm{eV}$.
\begin{figure}
\centering
\includegraphics[width=4.0in,angle=360]{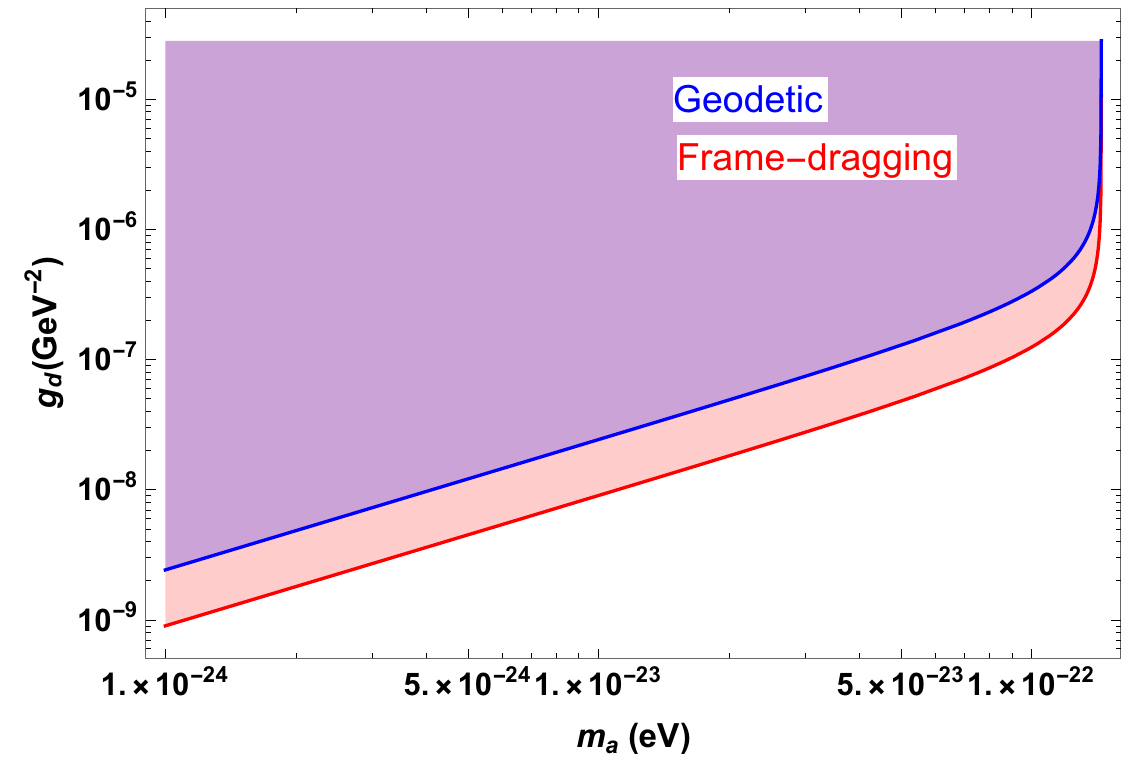}
\caption{Variation of $g_d$ with $m_a$ for axion EDM coupling from the geodetic (blue) and the frame-dragging (red) measurements using GP-B result. The shaded regions above these lines are excluded.}
\label{fig:plot5}
\end{figure}
In Fig.\ref{fig:plot5}, we have shown the variation of the coupling constant $g_d$ in $\rm{GeV^{-2}}$ for electric dipole moment operator with the axion mass $m_a$ in $\rm{eV}$. The red line denotes the variation of $g_d$ with respect to $m_a$ for frame dragging measurement whereas the blue line denotes the corresponding variation for the geodetic measurement. The shaded regions are excluded. Here the frame dragging gives the stronger bound. For $m_a\lesssim  1.45\times 10^{-22}~\rm{eV}$ the effects add coherently whereas for $m_a\gtrsim  1.45\times 10^{-22}~\rm{eV}$ the precessional velocity oscillates with time. For relatively heavier axions, one can constrain the axion EDM parameter from spin precession measurements, CMB spectra, and baryon acoustic oscillations as discussed in \cite{Graham:2013gfa,Budker:2013hfa,Caloni:2022uya}.
\section{Constraints on dark photon from GP-B result}
\label{sec6}
Suppose the background ultralight vector dark photon field $(A^\prime_\mu)$ couples with the nucleons $(N)$ of the gyroscope through electric and magnetic dipole moment operators described by the Lagrangian 
\begin{equation}
\mathcal{L}\supset g^\prime_{MDM}F^\prime_{\mu\nu}\bar{N}\sigma^{\mu\nu}N+g^\prime_{EDM}F^\prime_{\mu\nu}\bar{N}\gamma^5\sigma^{\mu\nu}N,
\label{k1}
\end{equation}
where the first term denotes the magnetic dipole moment operator, the second term denotes the electric dipole moment operator, and $F^\prime_{\mu\nu}=\partial_\mu A^\prime_\nu-\partial_\nu A^\prime_\mu$. In the frame of the gyroscope travelling with a velocity $\vec{v}$, the dark magnetic field is $\vec{B}_{A^\prime_\mu}=\vec{v}\times \vec{E}_{A^\prime_\mu}$, where $E_{A^\prime_\mu}\sim\sqrt{\rho_{DM}}$ is the amplitude of the oscillating dark electric field in the lab frame. Since, in the non relativistic limit $<F^\prime_{\mu\nu}\sigma^{\mu\nu}>\sim \vec{\sigma}_N.\vec{B}_{A^\prime_\mu}$ and $<F^\prime_{\mu\nu}\gamma^5\sigma^{\mu\nu}>\sim \vec{\sigma}_N.\vec{E}_{A^\prime_\mu}$, the precessional velocity of the gyroscope due to the magnetic dipole moment operator is 
\begin{equation}
v^{MDM}_{prec}(t)=g^\prime_{MDM}\sqrt{\rho_{DM}}v\cos\theta\cos(m_{A^\prime} t),
\label{k2}
\end{equation} 
where $\cos\theta$ picks the normal component of the dark electric field with respect to the gyroscopic plane and $m_{A^\prime}$ is the mass of the dark photon. Similarly, for the electric dipole moment operator, the precessional velocity due to the dark electric field is 
\begin{equation}
v^{EDM}_{prec}(t)=g^\prime_{EDM}\sqrt{\rho_{DM}}\sin\theta\cos(m_{A^\prime} t).
\label{k3}
\end{equation}
The background vector dark photon field changes the precessional velocity of the gyroscope through Eq.\ref{k2} and Eq.\ref{k3} and its contribution is within the uncertainties in the measurements of the precessional velocity of the gyroscope as obtained from GP-B result.
\begin{figure}
\centering
\includegraphics[width=4.0in,angle=360]{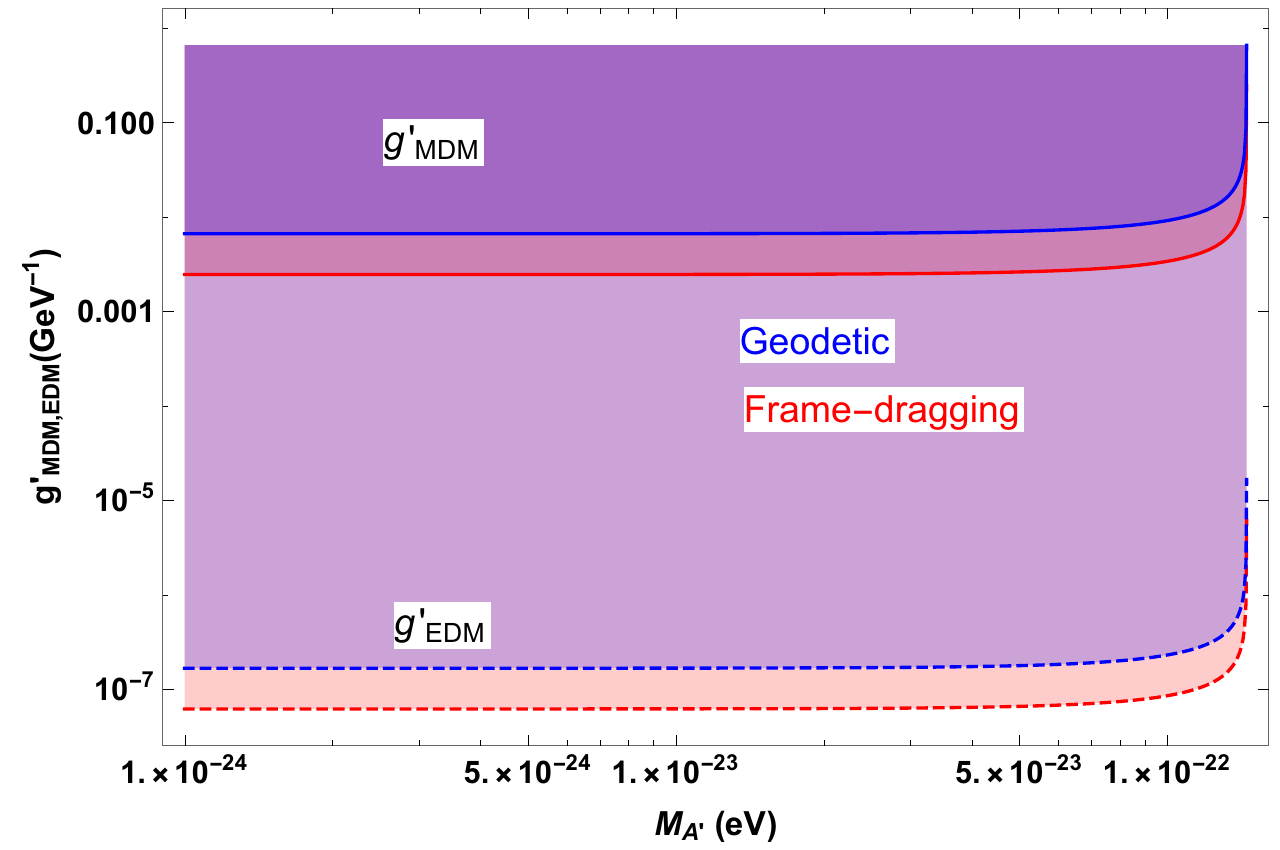}
\caption{Variation of dipole moment couplings $g^\prime_{MDM}$ and $g^\prime_{EDM}$ with the dark photon mass $m_{A^\prime}$ from the geodetic (blue) and the frame-dragging (red) measurements using GP-B result. The shaded regions above these lines are excluded.}
\label{fig:plot4}
\end{figure}
In Fig.\ref{fig:plot4}, we have shown the variation of $g^\prime_{MDM}$ and $g^\prime_{EDM}$ with the ultralight dark photon mass $m_{A^\prime}$ for the geodetic (blue line) and the frame-dragging (red line) measurements using GP-B result. Here, we have typically chosen $\cos\theta\sim 1$ and $\sin\theta\sim 1$. The dashed lines correspond to the EDM coupling whereas the solid lines correspond to the MDM coupling. The upper bounds on the MDM coupling from geodetic measurement is $ g^\prime_{MDM}\lesssim 6.7\times 10^{-3}~\rm{GeV^{-1}}$ whereas from frame dragging measurement, it is $ g^\prime_{MDM}\lesssim 2.5\times 10^{-3}~\rm{GeV^{-1}}$. We also obtain the upper bounds on the EDM coupling from geodetic measurement is $g^\prime_{EDM}\lesssim 1.82\times 10^{-7}~\rm{GeV^{-1}}$ whereas from frame dragging measurement, it is $g^\prime_{EDM}\lesssim 6.27\times 10^{-8}~\rm{GeV^{-1}}$. For both MDM and EDM, we obtain stronger bounds from frame dragging. However, EDM puts a stronger bound than MDM. The bounds on the EDM and MDM couplings are only valid for the dark photon of mass $m_{A^\prime}\lesssim \frac{2\pi}{T}\sim 1.45\times 10^{-22}~\rm{eV}$, where $T$ denotes the mission duration. 
\section{Constraints on vector gauge bosons in a gauged $L_e-L_{\mu,\tau}$ scenario}
\label{sec5}
Due to the presence of electrons inside the Earth and the gyroscope of the GP-B telescope, the exchange of $L_e-L_{\mu,\tau}$ gauge bosons between the electrons can give rise to a flavor dependent long range force between the satellite and the Earth. The long range force can contribute to the precession rate of the gyroscope and from the GP-B result, we obtain bounds on gauge boson coupling and its mass.

The electrons inside the Earth can generate a potential $V(r)$ at the quartz sphere surface as 
\begin{equation}
V(r)=\frac{g^2N}{4\pi r}e^{-M_{Z^\prime}r},
\label{eq:14}
\end{equation}
where $N\approx 3.35\times 10^{51}$ is the number of electrons inside the Earth, $M_{Z^\prime}$ is the mass of the gauge boson and $a=7027.4~\rm{km}=3.5\times 10^{13}~\rm{eV^{-1}}$ is the distance between the Earth and the gyroscope. The mass of the gauge boson is constrained by the inverse of the distance between Earth and the gyroscope which gives $M_{Z^\prime}\lesssim 2.82\times 10^{-14} ~\rm{eV}$. Hence, the electric field at the gyroscope due to the long range Yukawa potential is 
\begin{equation}
E_{Z^\prime}(r)=-\nabla V(r)=\frac{g^2N}{4\pi r}\Big(\frac{1}{r}+M_{Z^\prime}\Big)e^{-M_{Z^\prime}r}.
\label{eq:15}
\end{equation}
Now the orbital velocity of the satellite in the polar orbit is $v\sim 2.51\times 10^{-5}$. The magnetic field in the boosted frame tied with the GP-B satellite is $\vec{B}_{Z^\prime}\sim -\vec{v}\times \vec{E}_{Z^\prime}$ and its magnitude is
\begin{equation}
B_{Z^\prime}(r)=\frac{g^2 N v}{4\pi r}\Big(\frac{1}{r}+M_{Z^\prime}\Big)e^{-M_{Z^\prime}r}.
\label{eq:16}
\end{equation}
The radius of the superconducting quartz sphere gyroscope is $R=1.9~\rm{cm}=9.59\times 10^4~\rm{eV^{-1}}$ and the angular velocity is $\omega=9000~\rm{rpm}=6.20\times 10^{-13} ~\rm{rad.eV}$. The mass of the quartz sphere is $M=63~\rm{grams}=3.54\times 10^{25} ~\rm{GeV}$. The numerical value of the charge of the quartz sphere is the same as its mass and hence $q=3.54\times 10^{25}$. Since the gyroscope is rotating, so the magnetic moment of the rotating charged sphere is 
\begin{equation}
\mu=\frac{q}{5}R^2\omega=4.037\times 10^{22}~\rm{rad.eV^{-1}}.
\label{eq:17}
\end{equation}
Now the rate of change of angular momentum is $\frac{d\vec{L}}{dt}=\vec{\tau}=\vec{\mu}\times \vec{B}$ with $\frac{d\vec{L}}{dt}=\vec{\Omega}\times \vec{L}$, where $\hat{\vec{\Omega}}$ is the precession axis and $\Omega$ denotes the precession rate. For all the previous calculations, we have taken $L\sim \hbar=1$ since the particles were all microscopic fermions and that gives $\Omega=\mu B$. Now we have a macroscopic system for which $L=I\omega$, where $I\sim \frac{2}{5}MR^2$ is the moment of inertia of the sphere about its centre of mass.
Hence, the precessional velocity due to the exchange of $L_e-L_{\mu,\tau}$ gauge boson is 
\begin{equation}
\Omega=\frac{\mu\times B}{I\omega}=3.34\times 10^{36}\frac{g^2}{r}e^{-M_{Z^\prime}r}\Big(\frac{1}{r}+M_{Z^\prime}\Big)~\rm{rad.eV}.
\label{eq:18}
\end{equation}
\begin{figure}
\centering
\includegraphics[width=4.0in,angle=360]{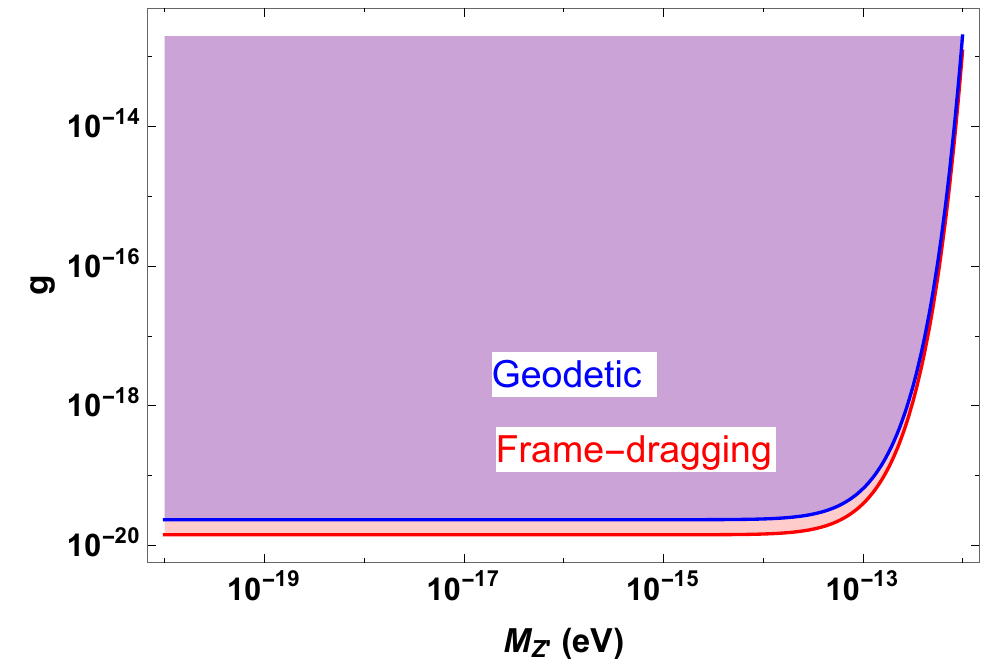}
\caption{Variation of $g$ with $M_{Z^\prime}$ for $L_e-L_{\mu,\tau}$ vector gauge boson mediation from the geodetic (blue) and the frame-dragging (red) measurements using GP-B result. The shaded regions above these lines are excluded.}
\label{fig:plot2}
\end{figure}
The contribution in gyroscope precession rate due to $L_e-L_{\mu,\tau}$ gauge bosons must be within the uncertainties that give bounds on $g$ as $g\lesssim 2.33\times 10^{-20}$ from geodetic measurement and $g\lesssim 1.42\times 10^{-20}$ from frame-dragging measurement. In Fig.\ref{fig:plot2}, we have shown the variation of gauge coupling with the gauge boson mass from geodetic (blue line) and frame-dragging measurements (red line). The regions above these lines are excluded. The frame-dragging puts the stronger bound on the gauge coupling for the gauge bosons of mass $M_{Z^\prime}\lesssim2.82\times 10^{-14}~\rm{eV}$. Though the bound on the gauge coupling is $10^5$ times weaker than that obtained from neutrino oscillation experiment \cite{Joshipura:2003jh} or the perihelion precession of planets \cite{Poddar:2020exe}, it is important to conclude the fact that such type of particle physics model can be constrained from GP-B experiment as well. Moreover, we suggest that the interaction between the gauge boson and the electron, giving rise to a long range fifth force in a gauged $L_e-L_{\mu,\tau}$ scenario can be constrained from geodetic and frame-dragging measurements from GP-B satellite that was initially built for gravity experiments.

\section{Constraints on unparticle mediated long range force from GP-B result}
\label{sec7}
The vector unparticle couples to the leptonic or the baryonic current through the effective coupling
\begin{equation}
 \mathcal{L}_{\rm{eff}} \supset \frac{c_u}{\Lambda_u^{d-1}}J^\mu \mathcal{O}^u_\mu,
\end{equation}
where $\mathcal{O}^u_\mu$ is the unparticle operator, $J_\mu$ is the baryonic or leptonic current, $c_u$ is the vector coupling, $\Lambda_u$ is the scale where the fields become conformally invariant, and $d$ is the scaling dimension. We assume that $\mathcal{O}^u$ and the fermion field $\psi$ obey the following $U(1)$ gauge symmetry
\begin{equation}
 \psi\rightarrow e^{i\alpha}\psi, \hspace{0.5cm} \mathcal{O}^u_\mu\rightarrow \mathcal{O}^u_\mu+\frac{\Lambda_u^{d-1}}{c_u}\partial_\mu \alpha.
\end{equation}
Hence, the unparticle remains massless below the scale $\Lambda_u$. One can write the unparticle propagator as
\begin{equation}
 D^{\mu\nu}=\frac{16\pi^{5/2}}{(2\pi)^{2d}}\frac{\Gamma(d+\frac{1}{2})}{\Gamma(d-1)\Gamma(2d)}K^{\mu\nu}(-k^2)^{d-2},
\end{equation}
where 
\begin{equation}
 K^{\mu\nu}(k)=\eta^{\mu\nu}-\frac{k^\mu k^\nu}{k^2}.
\end{equation}
By taking the Fourier transform of propagator in the static limit, we obtain the long range unparticle mediated potential at the surface of the quartz sphere due to the presence of SM particles in the Earth as 
\begin{equation}
V_u(r,d,\Lambda_u)=\frac{1}{2\pi^{2d}}c^2_u\Big(\frac{1}{\Lambda_u}\Big)^{2d-2}\frac{\Gamma(d+\frac{1}{2})\Gamma(d-\frac{1}{2})}{\Gamma(2d)}\frac{N}{r^{2d-1}},
\label{w1}
\end{equation}
where $N\approx 3.35\times 10^{51}$ is the number of electrons/nucleons in Earth. For the vector coupling, the forces add coherently. Hence, from Eq.\ref{w1} the electric field at the surface of the quartz sphere is 
\begin{equation}
E(r,d,\Lambda_u)=\frac{1}{2\pi^{2d}}c^2_u\Big(\frac{1}{\Lambda_u}\Big)^{2d-2}\frac{\Gamma(d+\frac{1}{2})\Gamma(d-\frac{1}{2})}{\Gamma(2d)}\frac{N}{r^{2d}}(2d-1).
\label{w2}
\end{equation} 
From Eq.\ref{w2}, we can write the corresponding magnetic field in the frame tied with the GP-B satellite is 
\begin{equation}
B(r,d,\Lambda_u)=\frac{1}{2\pi^{2d}}c^2_u\Big(\frac{1}{\Lambda_u}\Big)^{2d-2}\frac{\Gamma(d+\frac{1}{2})\Gamma(d-\frac{1}{2})}{\Gamma(2d)}\frac{N v}{r^{2d}}(2d-1),
\label{w3}
\end{equation}
where $v$ is the velocity of the satellite at the polar orbit which is $v\sim 2.51\times 10^{-5}$. We have previously calculated the magnetic moment of the rotating charged gyroscope as $\mu=4.037\times 10^{22}~\rm{rad.eV^{-1}}$. Hence, for the macroscopic system, the precessional velocity due to the exchange of unparticle is 
\begin{equation}
\Omega=\frac{\mu\times B}{I\omega}=4.198\times 10^{37}\frac{1}{2\pi^{2d}}c^2_u\Big(\frac{1}{\Lambda_u}\Big)^{2d-2}\frac{\Gamma(d+\frac{1}{2})\Gamma(d-\frac{1}{2})}{\Gamma(2d)}\frac{1}{r^{2d}}(2d-1).
\label{w4}
\end{equation}
The unparticle mediated long range force can contribute to the precessional velocity of the gyroscope and its contribution is within the experimental uncertainty in the measurements of geodetic and frame-dragging effects. Here, we have chosen $r$ as the Earth-satellite distance.
\begin{figure}
\centering
\includegraphics[width=4.0in,angle=360]{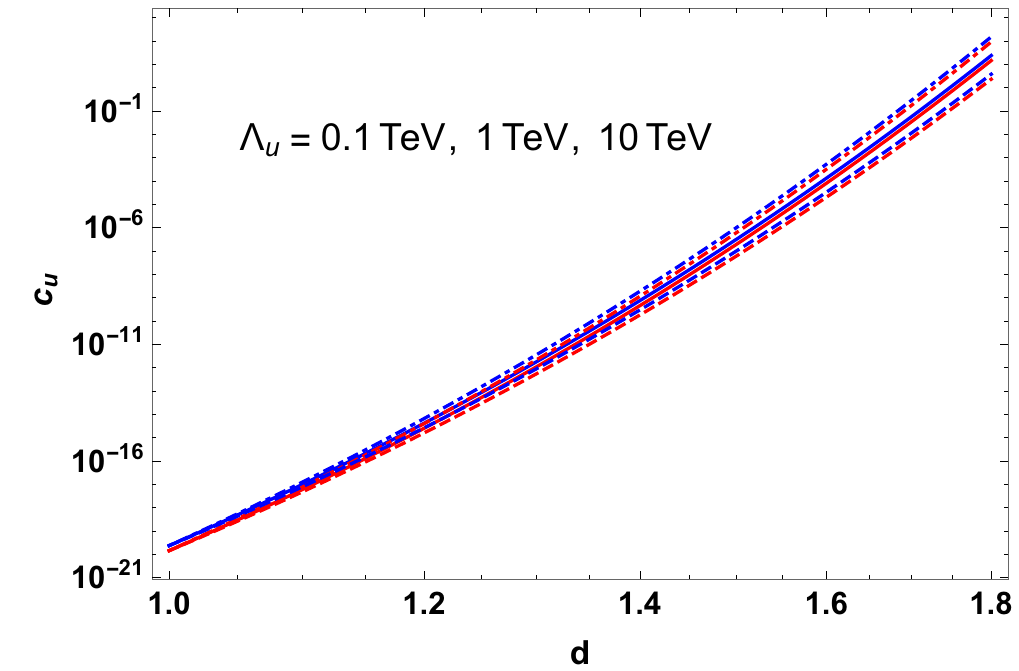}
\caption{Variation of vector coupling with the scaling dimension for unparticle exchange between the Earth and GP-B satellite from the geodetic (blue) and the frame-dragging (red) measurements. The solid, dot-dashed, and dashed lines correspond to the energy scale $1~\rm{TeV}$, $10~\rm{TeV}$, and $0.1~\rm{TeV}$ respectively. }
\label{fig:unparticle}
\end{figure}
In Fig.\ref{fig:unparticle}, we have shown the variation of vector coupling $c_u$ with the scaling dimension $d$ for unparticle exchange between the Earth and the GP-B satellite for the geodetic (blue lines) and frame-dragging (red lines) measurements. Here we have chosen the values of the energy scale as $\Lambda_u=1\hspace{0.1cm}\rm{TeV}$ (solid lines), $\Lambda_u=10\hspace{0.1cm}\rm{TeV}$ (dot-dashed lines) and $\Lambda_u=0.1\hspace{0.1cm}\rm{TeV}$ (dashed lines). With increasing $\Lambda_u$, $c_u$ increases. The frame-dragging effect gives the stronger bound on $c_u$ for any value of $d$. For $d=1$ we obtain $c_u\lesssim 1.83\times 10^{-20}$. The effect of the scale $\Lambda_u$ is prominent for larger values of $d$. This bound is complementary to that obtained from the perihelion precession of planets \cite{Das:2007cc}. 


\begin{table}
\caption{\label{tab1}Summary of bounds on spin-dependent and spin-independent couplings and masses of light particles from geodetic and frame dragging measurements.}
\begin{tabular}{ lcc  }
 \hline
Light particles & couplings & mass  \\
 \hline
Long range axions coupled with nucleons   &  $ f_a\gtrsim 1.6\times 10^{11}~\rm{GeV}$ & $\lesssim10^{-26}~\rm{eV}$\\
Oscillating axions coupled with nucleons & $ g_{aNN}\lesssim 1.7\times 10^{-3}~\rm{GeV^{-1}}$ & $\lesssim 1.45\times 10^{-22}~\rm{eV}$\\
Oscillating axions EDM coupling with nucleons   &  $ g_d\lesssim 9.4\times 10^{-10}~\rm{GeV^{-2}}$ & $\lesssim1.45\times 10^{-22}~\rm{eV}$\\
Oscillating dark photon EDM coupling with nucleons   &  $ g^\prime_{EDM}\lesssim 6.3\times 10^{-8}~\rm{GeV^{-1}}$ & $\lesssim 1.45\times 10^{-22}~\rm{eV}$\\
Oscillating dark photon MDM coupling with nucleons & $ g^\prime_{MDM}\lesssim 2.5\times 10^{-3}~\rm{GeV^{-1}}$ & $\lesssim1.45\times 10^{-22}~\rm{eV}$\\
$L_e-L_{\mu,\tau}$ long range gauge bosons & $ g\lesssim 1.4\times 10^{-20}$ & $\lesssim2.82\times 10^{-14}~\rm{eV}$\\
Unparticle mediated long range force  &  $ c_u\lesssim 1.8\times 10^{-20}$, $d=1$ & $0$\\
 \hline
\end{tabular} 
\end{table}
In TABLE \ref{tab1}, we have summarized the bounds on spin-dependent and spin-independent couplings of ultralight particles such as axions, gauge bosons, dark photons, and unparticles obtained from the GP-B results. We consider axial vector coupling, electric dipole moment coupling of axions with nucleons, electric and magnetic dipole moment couplings of the dark photons with nucleons, vector couplings of gauge boson, and unparticle with the SM particles. The bounds for axial vector and electric dipole moment coupling of axions, and electric and magnetic dipole moment coupling of the dark photon with nucleons are only valid for these ultralight particles of mass $\lesssim 10^{-22}~\rm{eV}$. For the spin independent vector coupling of $L_e-L_{\mu,\tau}$ gauge bosons, the bounds are only valid for gauge boson mass $\lesssim 10^{-14}~\rm{eV}$.
\section{Discussions}
\label{sec8}
In this paper, we consider a variety of new interactions that would possibly give rise to additional torques on the GP-B gyroscope. We use the GP-B results to place constraints on new physics parameters that parametrize these interactions. The new physics scenarios are axion mediated force between a CP violating dark sector (comprising the dark matter) and the SM, axion coupling with the nucleons through axial vector and EDM operators, dark photon coupling to the nucleon EDM and MDM operators, vector coupling of $L_e-L_{\mu,\tau}$ gauge bosons with the electrons in the GP-B, and vector unparticle coupling with nucleons in the GP-B gyroscope. We have obtained the bounds on the different operator couplings of ultralight axions, vector gauge bosons, and unparticles from the study of geodetic and frame-dragging measurements. The gyroscope of the GP-B satellite which measures the spacetime curvature near Earth contains lots of electrons and nucleons. Ultralight axions, vector gauge bosons, and unparticles can interact with those SM particles through different operator couplings and change the drift rate of the gyroscope. These ultralight particles can contribute to the precession rate of the gyroscope through different operator couplings and their contribution is within the uncertainties in the measurements of geodetic and frame-dragging effects. 

If the ultralight axions mediate a long range force between the dark sector and the GP-B gyroscope then the axion induced magnetic field can change the drift rate of the gyroscope and we obtain bound on the axion decay constant as $f_a\gtrsim 1.6\times 10^{11}~\rm{GeV}$ for the axions of mass $m_a\lesssim10^{-26}~\rm{eV}$. The ultralight axions can also behave as a background time oscillating field and when the gyroscope passes through this background field, the drift rate of the gyroscope changes. We obtain the axion nucleon coupling as $ g_{aNN}\lesssim 1.7\times 10^{-3}~\rm{GeV^{-1}}$ for the axions of mass $m_a\lesssim 1.45\times 10^{-22}~\rm{eV}$. The time oscillating background axion field can also couple with the nucleons of the gyroscope through electric dipole moment operator. The Earth's magnetic field induces the precession of the gyroscope and we obtain the upper bound on the coupling as $ g_d\lesssim 9.4\times 10^{-10}~\rm{GeV^{-2}}$ for axion mass $m_a\lesssim1.45\times 10^{-22}~\rm{eV}$.
 Time oscillating ultralight dark photon field can also interact with the nucleons of the gyroscope through electric and magnetic dipole moment operators and change the drift rate of the GP-B gyroscope. We obtain the upper bounds for EDM operator as $g^\prime_{EDM}\lesssim 6.3\times 10^{-8}~\rm{GeV^{-1}}$ and for MDM operator as $ g^\prime_{MDM}\lesssim 2.5\times 10^{-3}~\rm{GeV^{-1}}$ for the dark photon of mass $m_{A^\prime}\lesssim1.45\times 10^{-22}~\rm{eV}$. We obtain a stronger bound for EDM coupling than the MDM coupling. Due to the presence of electrons in the GP-B gyroscope and the Earth, long range force of $L_e-L_{\mu,\tau}$ type can mediate between the Earth and the GP-B satellite. The mediation of $L_e-L_{\mu,\tau}$ gauge bosons can change the drift rate of the gyroscope and we obtain the bound on the gauge coupling as $g\lesssim 1.4\times 10^{-20}$ for ultralight vector gauge boson mass $M_{Z^\prime}\lesssim2.82\times 10^{-14}~\rm{eV}$. The massless unparticles can also mediate long range force between the Earth and the gyroscope and alter the drift rate of the GP-B satellite. We obtain the bound on the unparticle coupling as $c_u\lesssim 1.8\times 10^{-20}$ for the scaling dimension $d=1$. The frame-dragging puts stronger bounds on all the above mentioned operator couplings. The bounds for axial vector and electric dipole moment coupling of axions, and electric and magnetic dipole moment coupling of the dark photon with nucleons are only valid for these ultralight particles of mass $\lesssim 10^{-22}~\rm{eV}$. For the spin independent vector coupling of $L_e-L_{\mu,\tau}$ gauge bosons, the bounds are only valid for gauge boson mass $\lesssim 10^{-14}~\rm{eV}$.
 
 For the spin-independent vector couplings of ultralight particles, the forces add coherently. However, for spin-dependent operator couplings, a fraction of spins of SM particles in the gyroscope needs to be polarized. Here, we show a small fraction of polarization exists in the gyroscope due to the London-induced magnetic moment. The spin-dependent long range forces have an explicit dependence on the spin polarization fraction in the gyroscope. Alternatively, instead of a quartz sphere, one can make spin polarized ferromagnetic gyroscope or single domain magnetic needle \cite{Fadeev:2020gjk,PhysRevLett.116.190801} in the laboratory to obtain stronger bounds for the spin dependent couplings of axions, dark photons, and unparticles. Such a laboratory made spin polarized gyroscope can be used in GP-B satellite (future mission, if any) and one can constrain spin dependent long range forces for astrophysical distances which was made to test Einstein's theory of gravity. Other than GP-B, laser ranging network \cite{Lucchesi:2018cdl,PEARLMAN2002135,Ciufolini:2016ntr,universe4110113,doi:10.1126/science.279.5359.2100}, interferometry method \cite{PhysRevResearch.2.032069,DIVIRGILIO2014866} can also measure the geodetic and frame-dragging effects. Future experiments like LARES2 \cite{Ciufolini:2017tnj,Ciufolini:2021fon} which can measure the frame-dragging with better sensitivity can significantly improve our bounds.

\section*{Acknowledgements}
The author would like to thank Subhendra Mohanty for his valuable comments and suggestions. The author is also grateful to Surjeet Rajendran, Dongok Kim, Younggeun Kim, Yannis K. Semertzidis, Yun Chang Shin, Wen Yin for useful discussions. The author is also thankful to the anonymous referee for useful comments and suggestions.
\bibliographystyle{utphys}
\bibliography{cosmicaxion}
\end{document}